\documentclass[]{aa}
\usepackage{txfonts}
\usepackage[T1]{fontenc}
\usepackage{amssymb}
\usepackage{natbib}
\usepackage{bm}
\usepackage{palatino}
\usepackage{graphicx,color,epsf}

\newcommand{\ud}{\mathrm{d}}

\begin{document}

\title{Radiation pressure and pulsation effects on the Roche lobe}
\titlerunning{Radiation pressure and pulsation effects on the Roche lobe}
\author{T.~Dermine\inst{1} \and A.~Jorissen\inst{1} \and L.~Siess\inst{1}
  \and A.~Frankowski\inst{2}}
\institute
    {Institut d'Astronomie et d'Astrophysique, Universit\'e
      libre de Bruxelles, Facult\'e des Sciences, CP. 226, Boulevard du
      Triomphe, B-1050 Bruxelles, Belgium
      \and Department of Physics, Technion-Israel Institute of Technology,
      32000 Haifa, Israel
    }
    
\date{Received date; accepted date} 

\abstract
    {Several observational pieces of evidence indicate that 
      specific evolutionary channels which involve Roche lobe overflow are not
      correctly accounted for by the classical Roche model.}
    {We generalize the concept of Roche lobe in the presence of
      extra forces (caused by radiation pressure or pulsations). 
      By computing the distortion of the equipotential surfaces, we
      are able to evaluate the impact of   these perturbing forces
      on the stability of Roche-lobe overflow (RLOF).
    }
    {Radiative forces are parameterized through the constant reduction factor
      that they impose on the gravitational force from the radiating star (neglecting any
      shielding in case of  large optical thickness). Forces imparted by pulsations
      are derived from the velocity profile of the wind that they trigger.
    }
    {We provide analytical expressions to compute the generalized Roche radius.
      Depending on the extra force, the Roche-lobe radius may either stay
      unchanged, become smaller, or even become meaningless (in the presence of
      a radiatively- or pulsation-driven wind). There is little
      impact on the RLOF stability.}
    {}

\keywords{binaries: general - Stars: mass loss - Stars: winds, outflows}
\maketitle

\section{Introduction}
\label{Sect:Intro}

The Roche model has been widely used to infer the outcome of binary star
evolution. In this model, only the gravitational and centrifugal
forces are accounted for to compute the equipotential surfaces. However
if other forces are present in the system, like those responsible for mass
loss, they should be included in the description as well. The
modification of the usual Roche model was first pointed out by
\cite{Schuerman-1972} in the context of binary systems involving
early-type main sequence stars with a strong wind.
The idea was further explored in various directions
by \citet{Kondo-1976}, \citet{Vanbeveren-1977,Vanbeveren-1978}, \citet{Friend-1982},
  \citet{Djurasevic-1986}, \citet{Zhou-1988}, \citet{Huang-1990}, \citet{Drechsel-1995}, 
  \citet{Frankowski-2001}, \citet{Phillips-2002} and \citet{Owocki-2007}.
In the present paper, we investigate in a more systematic way the
situations where the Roche model should be modified by considering the
different physical processes driving the stellar winds, that we briefly
describe in Sect.~\ref{Sect:Wind mass loss}.

The Roche model is generally used to answer two different questions,
namely: (i)  What is the flow geometry? (ii) Is the star filling its
Roche lobe? The first question is related to the
geometry of the equipotentials (and to the Coriolis force). One
important modification of the
equipotential geometry which can arise in the presence of an extra force
pervading all space (like radiation pressure), is that the
equipotentials open up in the direction of the external Lagrangian
point, thus possibly allowing the matter ejected
by the mass-losing component to form a circumbinary disc. 
That issue is important in the framework of the binary
evolution involving low- and intermediate-mass components  where the
formation of a circumbinary disc seems to be very common
\citep[see][]{DeRuyter-2006,Frankowski-2007a,Frankowski-2009a}.

The second question corresponds to the Roche-lobe overflow (RLOF)
criterion, which involves the comparison of the Roche radius with the
stellar radius. In the present paper we show how, depending on the physical
process driving the wind, the Roche radius may either be unchanged with
respect to the classical Roche model, become smaller, or even become
meaningless. This will depend on the value of the extra force at the
stellar surface (in contrast to the first question,
an answer to which requires the knowledge of the force everywhere within
the system). In a mass-losing star, the photospheric radius itself
  may become ill-defined, thus complicating the use of the RLOF criterion (see Sect.~\ref{Sect:Mass losing stars}).
 
Both issues (the equipotential geometry and the RLOF criterion) will be
addressed in the present paper for the generalised Roche model when
radiation pressure or pulsations play a role (Sect.~\ref{Sect:The effective
  potential}).  The way to correctly account for radiation pressure is
discussed in Sect.~\ref{Sect:radiation}. Typical values for the radiation pressure at the surface of various classes of stars are given in Sect.~\ref{Sect:Determination of f values}, and the corresponding shapes of the equipotentials are displayed in Sect.~\ref{Sect:Roche1}.
A numerical fit to the Roche radius will be provided in
Sect.~\ref{Sect:Roche}, generalising Eggleton's formula
\cite[]{Eggleton-83} to situations where a radiation-pressure force is
present. The necessity to abandon the Roche-lobe concept in the case of
stars suffering from radiatively- or pulsation-driven
wind mass loss is demonstrated in Sect.~\ref{Sect:Mass losing stars}. Conclusions are drawn in Sect.~\ref{Sect:Conclusions}.

\section{The different modes of wind mass loss}
\label{Sect:Wind mass loss}

\cite{Holzer-1985}, \cite{Schatzman-93}, \cite{Lamers-1997},
\cite{Willson-00} and \citet{Owocki-2004} have reviewed the different types
of winds existing across the Hertzsprung-Russell diagram according to their
respective driving mechanisms. These mechanisms may be grouped in three
broad classes: radiation-driven winds (associated with high-luminosity
objects), pulsation-initiated winds, and Alfv\'en wave-induced winds.

The first class includes line-driven winds, operating in {\em luminous hot
  stars} (OB stars, and Wolf-Rayet --WR-- stars as well) on resonance and
subordinate lines \citep{Castor-75,Abbott-1982}. Fast winds with terminal
velocities of the order of $500-3000 \textrm{ km s}^{-1}$ are generated
\citep[]{Kudritzki-00}. At the same luminosities as O stars, WR stars have
larger mass-loss rates, so another mechanism -- like pulsation-driven mass
loss, or multiple scattering of photons in an optically-thick wind with an
ionisation stratification -- is probably adding to radiation pressure on
atomic lines
\citep{Glatzel-93,Owocki-99,Nugis-2000,Owocki-2004}. Dust-driven winds
belong to the same category as radiation-driven winds, but operate instead
in {\em luminous cool stars} on the asymptotic giant branch and require
wind densities large enough to couple dust with gas
\citep{Gail-87a,Lamers-1997,Wachter-02,Schroder-03,Sandin-2003,Sandin-2008}.
Wind terminal velocities are small, of the order of $10-15 \textrm{ km
  s}^{-1}$. An important specificity of cool-star winds is that the driving 
  force only slightly exceeds the gravitational attraction, as apparent from 
  their small terminal velocities.

The dust formation requires another process to lift the matter high enough
above the photosphere, in the region where the temperature is less than
1500 K. This can be done through shock waves associated with stellar
pulsation \citep{Bowen-88}, like in {\em Mira or semi-regular variables}.

Alfv\'en waves are the other main class of waves driving winds in stars
with open magnetic field lines. This latter wind mechanism is important for
stars that are not luminous enough to have a strong radiation pressure,
i.e., {\em magnetic A, F, G, K and M stars with luminosities lower than
  about $50$~L$_\odot$} (see Sect.~\ref{Sect:Determination of f values}).
It is the leading candidate to account for the solar wind. This driving
process has the important property to be a local phenomenon which does not
derive from a potential contrarily to radiation pressure.

\section{The effective potential}
\label{Sect:The effective potential}

In the Roche model, the two components of a binary system are considered as
point sources in circular orbits and in synchronous rotation with the
orbital motion. It is then possible to define a reference frame in uniform
rotation about the centre of mass of the system, in which the two stars are
at rest. When distances are expressed in units of the orbital separation,
time in units of the orbital period and masses in units of the total mass
($M_1+M_2$), the effective potential of the system, including the
gravitational and the centrifugal potentials, is given by
\begin{equation}
  \label{Eq:Roche}
  \Phi=-\frac{\mu}{r_1}-\frac{1-\mu}{r_2}-\frac{x^2+y^2}{2}
\end{equation}
where
\begin{displaymath}
  \begin{array}{ll}
    \mu=\frac{M_1}{M_1+M_2}
  \end{array}
\end{displaymath}
and
\begin{displaymath}
  \begin{array}{l}
    r_1=\left((x+1-\mu)^2+y^2+z^2\right)^{1/2}\\
    r_2=\left((x-\mu)^2+y^2+z^2\right)^{1/2}\\
  \end{array}
\end{displaymath}
are respectively the distance of a test particle located at $(x,y,z)$ to
the primary star (labelled "1" and located at $x_1=\mu-1$) and to the
companion (labelled "2" and located at $x_2=\mu$). The centre of mass is
located at the origin of the coordinate system. The mass of the test
particle is supposed to be small enough not to disturb the potential.
When additional forces deriving from a potential are present in the system
(other than the gravitational and centrifugal ones), the effective
potential writes
\begin{displaymath}
  \label{Eq:Phiextra}
  \Phi=\Phi_{\rm extra}-\frac{\mu}{r_1}-\frac{1-\mu}{r_2}-\frac{x^2+y^2}{2}.
\end{displaymath}

When the extra force is caused by radiation pressure, $\Phi_{\rm extra}$ is
easy to evaluate because the radiation force has the same $1/r^2$ dependence on
the distance as the gravitational attraction and it also
pervades all space. The validity of this simple model is further discussed
in Sect.~\ref{Sect:radiation}. The ratio $f$ of the radiation to the
gravitational force of star~1
\begin{equation}
  f\equiv -\frac{1}{\rho} \frac{\ud P_{\mathrm{rad}}}{\ud r}_{\rm r_1} \left( \frac{G
    M_1}{r_1^2}\right)^{-1} \mathrm{}
\end{equation}
is independent of $r_1$ provided that the radiation flux at position $r_1$ emanating 
from star 1
-- denoted $F_{\nu,1}(r_1)$ -- and appearing in the expression
\begin{equation}
  \label{Eq:Prad}
  \frac{{\rm  d}P_{\mathrm{rad}}}{{\rm  d}r}_{\rm r_1} = -\frac{1}{c}
  \int_0^\infty \kappa_\nu\; \rho\; F_{\nu,1}(r_1)\; {\rm 
    d}\nu,
\end{equation}
may be replaced by $L_{\nu,1}/(4\pi r_1^2)$, so that 
\begin{equation}
  f = \frac{1}{4\pi c GM_1} \int_0^\infty \kappa_\nu\; L_{\nu,1}\;
  {\rm  d}\nu\ .
  \label{Eq:f}
\end{equation}
In the above equations, 
$\rho$ is the density of the circumstellar matter, $\kappa_{\nu}$ is the
absorption coefficient per unit mass at frequency $\nu$, $c$ the speed of
light, and $L_{\nu,1}$ the luminosity of star 1 in the frequency range
$(\nu, \nu+$d$\nu$).
We will ignore radiation shielding and assume that the medium is
optically thin \citep[see also][]{Huang-1990,Drechsel-1995}.
If radiation does not reach thermal equilibrium with circumstellar matter (i.e., the matter is optically thin), the radiation luminosity remains 
approximately constant all over the place, and the radiation flux 
can be expressed as $F_{\nu,1}(r_1)=L_{\nu,1}/4\pi r_1^2$, 
which may be rewritten $F_{\nu,1}(r_1)=F^*_{\nu,1}(R_1/r_1)^2$, $F^*_{\nu,1}$ being the radiation flux emitted per unit surface by the star of radius $R_1$. If that star is tidally distorted, limb- and gravity-darkened, a rigourous treatment would imply that Eq.~\ref{Eq:Prad} depends on the spherical coordinates $\theta, \phi$ since the flux irradiated by star 1 in a given direction defined by $\theta, \phi$ will now depend on these variables. 
The inclusion of these complications is beyond the scope of this paper, but they are  not expected to alter our general conclusions. 
It is assumed as well that $\kappa_\nu$ does not depend strongly on the optical
depth $\tau$ in the wind, i.e. that the ionisation state and/or level
population of the gas do not change. This issue has been thoroughly studied
in the context of radiatively-driven hot-star winds by \citet{Abbott-1982},
\citet{Pauldrach-1986}, \citet{Shimada-1994}, \citet{Gayley-1995},
\citet{Lamers-1997} and \citet{Puls-2000}. We thus consider here an idealised situation where $f$ is assumed constant. Under these hypotheses, the
effective gravity of the mass-losing star is then reduced by a factor
$(1-f)$, the same everywhere, and the effective potential becomes
\begin{equation}
  \label{Eq:phi_gen}
  \Phi=-\frac{\mu (1-f)}{r_1}-\frac{1-\mu}{r_2}-\frac{x^2+y^2}{2}\textrm{\
    .}
\end{equation}

However, even though the equipotentials give an idea of the flow geometry,
its precise description is only given by the equation of motion:
\begin{equation}
  \label{Eq:mvt}
  \frac{\ud^2 \mathbf{r}}{\ud t^2}=-\frac{1}{\rho} \nabla
  P-2\mathbf{\Omega}\times\frac{\ud \mathbf{r}}{\ud t}-\nabla
  \Phi+\mathbf{F}'\textrm{,}
\end{equation}
which includes the important Coriolis term. In the above expression, $P$ is
the gas pressure (this term can be important in case of very dense winds
like in Wolf-Rayet stars), $\mathbf{\Omega}$ is the angular velocity,
$\mathbf{F}'$ is the extra force and $\Phi$ is the Roche potential
(Eq.~\ref{Eq:Roche}).  This equation of motion will not be used in the
present paper.

\section{How to correctly account for radiation pressure?}
\label{Sect:radiation}

The impact of radiation pressure on the shape of the equipotential surfaces
has been discussed by several authors
\citep{Schuerman-1972,Huang-1990,Drechsel-1995,Howarth-1997,Maeder-Meynet-2000,Howarth-2001,Phillips-2002,Owocki-2007},
in slightly different contexts, and there has been some controversy as to
whether or not this effect modifies the Roche geometry. To make the
discussion clear, one should distinguish three different situations: (i)
the effect of radiation pressure on the matter outside the stars and
flowing in the binary system; (ii) the impact of the external radiation
field (from the companion star) on the equilibrium shape of the irradiated
star and (iii) the effect of the star's own radiation pressure on its
equilibrium configuration.

The first effect is relevant for deriving the flow geometry, and the
radiation pressure must always be included in the
computation. Complications here are only related to possible shielding
effects. In the present paper, we consider the situation where radiation
pressure comes only from the primary and neglect any shielding by the
companion star or by the circumstellar material which remains optically
thin everywhere.  Case (ii) has been considered by \citet{Drechsel-1995}, 
\citet{Phillips-2002} and \citet{Owocki-2007} 
in full details. In the present approach, it is
not accounted for and would require the incorporation of a new ($1-f'$)
factor reducing the gravity of the star labelled~2 in Eq.~\ref{Eq:phi_gen}.
\citet{Drechsel-1995} and \citet{Phillips-2002} have performed detailed
numerical calculations for this situation, including geometrical shielding
effects (not considered in our simple analytical approach) but the impact
is usually small. Case (iii) is conceptually more intricate and over the
years has become controversial
\citep{Friend-1982,Howarth-1997,Maeder-Meynet-2000,Howarth-2001}.  To
understand why, one should first distinguish the optically-thick from the
optically-thin regimes, following the insightful discussion of
\citet{Friend-1982}. In optically-thick regions, the diffusion
approximation for the radiative transfer may be used, resulting in an {\em
  isotropic} pressure with the (thermal equilibrium) value $(1/3) aT^4$.
Therefore, in the {\em hydrostatic} layers below the stellar photosphere
where this regime holds (this excludes WR stars), the effect of radiation
should be included in the pressure term
in the hydrostatic equilibrium equation and the
gravitational potential is unmodified. On the other hand, in an
optically-thin region, the radiation field is partially decoupled from the
gas and the stellar flux can be treated in a free-streaming
approximation. In this regime, the radiative flux is radial and
proportional to $1/r^2$ and acts as a repulsive force that reduces the
gravitational force by a factor $(1-f)$.

The diffusion and streaming approximations hold respectively deep within the star and in the outer layers of wind. Near the photosphere or in the surface layers where the hydrostatic equilibrium condition breaks down (like in WR stars, see Sect.~\ref{Sect:Mass losing stars}) the situation is more ambiguous as the von Zeipel's theorem does not apply \citep{Howarth-1997}.

\begin{figure}
  \centering
  \includegraphics[width=0.45\textwidth]{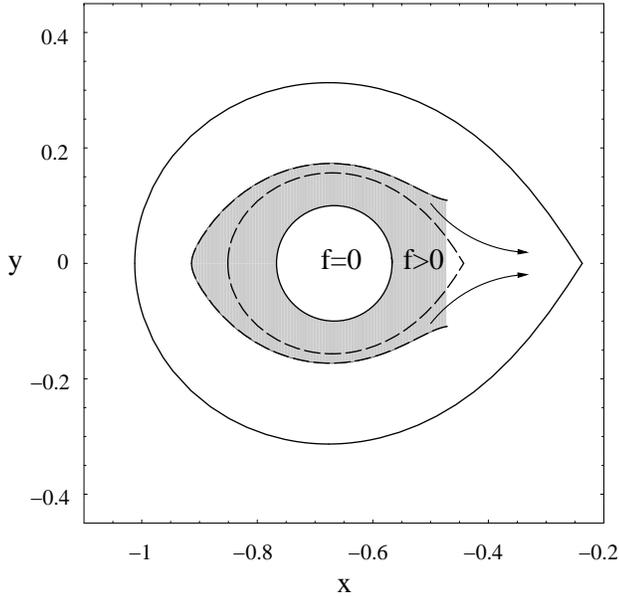}
  \caption{\label{Fig:dichotomy} A giant star with an extended atmosphere
    (grey region, on an exaggerated scale). As discussed in the text,
    in the optically-thin ($\tau < 2/3$) layers of the atmosphere, the
    ratio $f$ of the radiative force to gravitational attraction is
    positive and affects the Roche potential (dashed
    lines). Within the photosphere, the matter is optically thick, so that
    $f = 0$ and the classical Roche potential applies (solid lines). The
    part of the atmosphere located above its critical Roche lobe will flow
    through the companion (depicted by the arrows).  }
\end{figure}

To illustrate this dichotomy, consider
Fig.~\ref{Fig:dichotomy} which shows the Roche-lobe filling criterion in a
giant star at ($f=0$) and above ($f> 0$) the photosphere where the
radiation pressure modifies the potential. The unmodified Roche potential
(Eq.~\ref{Eq:Roche}) should be used for the subphotospheric matter, whereas
the modified potential (Eq.~\ref{Eq:phi_gen}) should be used above the
photosphere (grey area on Fig.~\ref{Fig:dichotomy}). Since the modified
potential leads to a smaller critical surface (as will be shown in
Sect.~\ref{Sect:Roche}), the situation may arise where the photosphere does
not fill its Roche lobe, but the supra-photospheric matter does, especially
for (super) giant stars with extended atmospheres. These supra-photospheric
shells are removed from the star, and in the case of a giant star with a
convective envelope, this removal will cause the photosphere to expand on a
thermal time scale \citep{Ritter-1996}. Therefore, even though the {\em
  photosphere} itself does not fill its own Roche lobe, the consideration
of the modified Roche potential alters the evolution, {\em at least for
  giant stars}. This distinction between dwarfs and giants is important,
because empirical arguments call for the use of a modified Roche
equipotential for giants, but not for dwarfs.  \citet{Howarth-1997} has
given convincing arguments that the {\em unmodified} potential should be
used whenever there is evidence for gravity darkening (a direct consequence
of von Zeipel's theorem), as is the case for dwarf stars in binary systems
\citep{Anderson-1977,Rafert-1980}. The situation is however different for
giants: an important empirical motivation for considering the case $f >
0$ is that it may account for the fact that some giant
stars exhibit ellipsoidal variations (due to non-sphericity) despite small
(classical) Roche-filling factors. Those ellipsoidal variables would appear
quite enigmatic without the presence of an additional force reducing the
actual Roche radius and disturbing the stellar shape.  As it will be shown
in Sect.~\ref{Sect:Roche}, the modified Roche radius (for $f > 0$) is
smaller than the classical one (for $f = 0$). Therefore, stars nearly
filling their Roche lobe with $f > 0$ would be far from filling their lobe
if that filling factor were estimated with the traditional formula
corresponding to $f = 0$. This is probably what happens for the $13$ s-type
symbiotic systems with ellipsoidal variations detected by
\cite{Mikolajewska-2007}, which have a (classical) Roche filling factor of
only $0.4$ -- $0.5$. Based on the analysis of the orbital circularisation
in a sample of binary systems with M-giant primaries, 
\citet{Frankowski-2009} find that these giants also do not fill more
than $\sim 0.5$ of their classical Roche lobes. Again, the explanation
may lie in a decrease in the Roche radius
below its classical value.

\section{Typical $\bm{f}$ values}
\label{Sect:Determination of f values}

To clarify the ideas, we now evaluate the $f$ factor using the
Castor-Abbott-Klein \citep[][hereafter CAK]{Castor-75} theory of
line-driven winds.  Let $g_{\rm grav} (r) = \frac{G\;M}{r^2}$ be the
gravitational acceleration at distance $r$ from the stellar centre, and
\begin{equation}
  g_{\rm rad} \equiv g_{\rm Th} (1 + {\cal F}(\tau)) = \frac{\kappa_e
    L}{4\pi\;c\;r^2} (1 + {\cal F}(\tau)),
\end{equation}
be the sum of the radiative accelerations due to Thomson scattering
($g_{\rm Th}$) and to an ensemble of lines ($g_{\rm Th}\; {\cal F}(\tau)$),
where $\cal F$ is the so-called "force multiplier". In the above
relation, $\kappa_e = \sigma_e \; n_e / \rho$ is the opacity coefficient
per unit mass for Thomson scattering, $\sigma_e$ is the corresponding
Thomson cross section ($6.65\; 10^{-25}$~cm$^{-2}$), $n_e$ is the electron
number density and $\rho$ is the density of the wind. If the wind is fully
ionised, $\kappa_e = \sigma_e \; \frac{1}{m_H} \left< \frac{Z}{A}\right>
\sim 0.2 (1 + X)$~cm$^{2}$~g$^{-1}$, where $m_H$ is the hydrogen mass. The
force multiplier has been computed by CAK, \citet{Abbott-1982},
\citet{Pauldrach-1986}, \citet{Shimada-1994}, \citet{Gayley-1995},
\citet{Lamers-1997} and \citet{Puls-2000} for various wind
thermodynamical conditions.  It turns out to be very small at optical
depths close to one because of the "saturation effect": radiation with
frequencies matching those of atomic transitions are efficiently removed
from the flux at the base of the photosphere and no flux at those
frequencies is left further up in a static atmosphere to accelerate
matter. Of course, the situation changes in the presence of a wind, since
the associated velocity gradient will then induce a Doppler shift of the
line frequencies. At low optical depths (i.e., higher up in the
expanding atmosphere), when all lines become optically thin, ${\cal
  F}(\tau)$ becomes very large, and reaches asymptotic values up to about
2000 \citep[see e.g., Table 2 of ][also Gayley 1995]{Abbott-1982}.

Using the Eddington parameter 
\begin{equation}
  \label{Eq:Gamma}
  \Gamma  = \frac{\kappa_e\; L}{4\pi\;c\;G\;M}
\end{equation} 
we find that 
\begin{equation}
  f = \Gamma \; (1+\cal{F})\mathrm{.}
\end{equation}
In Table~\ref{Tab:f}, we have listed the $f$ and $\Gamma$ values for stars
along the main sequence. The numbers are obtained by considering an
optically-thin medium, where $\cal F$ reaches its maximum value of about
2000 \citep{Abbott-1982,Gayley-1995}.

\begin{table}
  \begin{center}
    \caption{\label{Tab:f}
      The Eddington parameter $\Gamma$ (Eq.~\ref{Eq:Gamma}) for stars along the
      main sequence, adopting $\kappa_e =
      0.35$~cm$^{2}$~g$^{-1}$. Since ${\cal F} < 2000$ \citep{Abbott-1982},  $f_{\rm max} = 2000
      \Gamma$. Luminosities and masses are from \cite{Cox-2000}.
    }
    \begin{tabular}{llllll}
      \hline\hline
      Sp. Typ. & $\log$ ($L/L_\odot)$  & $M$  & $\Gamma$ & $f_{\rm max}$\cr
      & & (M$_\odot$)\cr
      \hline
      O5V & 5.95 & 60.0  & $4.0\;10^{-1}$ & 807.\cr
      B0V & 4.77 & 17.5  & $9.2\;10^{-2}$ & 184.\cr
      B5V & 2.97 & 5.9   & $4.3\;10^{-3}$ & 8.68\cr
      B8V & 2.33 & 3.8   & $1.5\;10^{-3}$ & 3.06\cr
      A0V & 1.77 & 2.9   & $5.5\;10^{-4}$ & 1.10\cr
      A5V & 1.19 & 2.0   & $2.1\;10^{-4}$ & 0.42\cr
      F0V & 0.86 & 1.6   & $1.3\;10^{-4}$ & 0.25\cr
      F5V & 0.51 & 1.4   & $6.4\;10^{-5}$ & 0.13\cr
      \hline
    \end{tabular}
  \end{center}
\end{table}
\begin{table}
  \begin{center}
    \caption{\label{Tab:fRGB}
      The $f$ parameter for \mbox{$M=1 \mathrm{M}_\odot$} K and M giants and subgiants with C/O$=0.5$, at $\tau_{\rm Ross}=1$.
    }
    \begin{tabular}{ccc}
      \hline\hline
      $T_{\rm eff}$ (K) & $\log g$ & $f$\cr
      \hline
      5000 & 0 & 0.97\cr
      4000 & 0 & 0.43\cr
      3000 & 0 & 0.18\cr
      5000 & 3 & 9.6 10$^{-4}$\cr
      4000 & 3 & 4.3 10$^{-4}$\cr
      3000 & 3 & 1.2 10$^{-4}$\cr
      \hline
    \end{tabular}
  \end{center}
\end{table}
The threshold $f = 1$ is found around spectral type A0 (corresponding to
stars with luminosities less than about $50$ L$_\odot$), and this result is
consistent with empirical arguments stating that the peculiarities of Am
stars cannot survive if the mass loss is too strong
\citep{Michaud-1983,Michaud-1986,Lemke-1990,Babel-1992}.  Furthermore, the
very existence of Am stars demonstrates the importance of the radiative
acceleration on the distribution of various chemical elements of these
stars \citep[e.g.][]{HuiBonHoa-2002,Alecian-2002}.

For late-type stars, the CAK model cannot be used because this
model does not include the relevant transitions, especially  molecular lines which may 
contribute significantly to the radiative driving force \citep{Jorgensen-1992}. 
Therefore, radiative accelerations for late-type stars have been taken  
from the MARCS model atmospheres \citep{Gustafsson-2008} for K and M giants. 
For those stars, the radiation pressure is dominated by the contribution of the near-infrared continuum where these stars emit most of their radiation. 
The corresponding $f$ values are listed in Table~\ref{Tab:fRGB}.


For AGB stars with still higher luminosities than those considered in Table~\ref{Tab:fRGB}, the radiation driving force
is now dominated by absorption and scattering in molecular lines, yielding values
of $f$ as large as $0.15$ for C-type stars with C/O~$=2$, $\log g = -1$
and $T_{\rm{eff}}=2500$~K \citep{Elitzur-1989,Jorgensen-1992}.

In the next sections, the values of $f$ quoted above will be applied to
different astrophysical situations.  It has to be made very clear that
one should distinguish situations involving RLOF (Sect.~\ref{Sect:RLOF})
from situations involving modifications of the geometry of the
equipotential surfaces far above the photosphere of the mass-losing star
(Sect.~\ref{Sect:Roche1}).  Although in the latter case, the $f$ values
provided in Table~\ref{Tab:f} may be used without restrictions, the
situation is a bit more complicated in the former case. This is because
the RLOF criterion involves photospheric layers, and we have argued in
Sect.~\ref{Sect:radiation} that non-zero $f$ values associated with
radiation pressure {\em do not usually alter} the stellar equilibrium
configuration (according to the von Zeipel theorem), except in special
circumstances involving giant stars. The two situations are therefore
discussed in separate sections below.

\section{The modified Roche equipotentials}
\label{Sect:Roche1}
\begin{figure}
  \centering
  \includegraphics[width=0.45\textwidth]{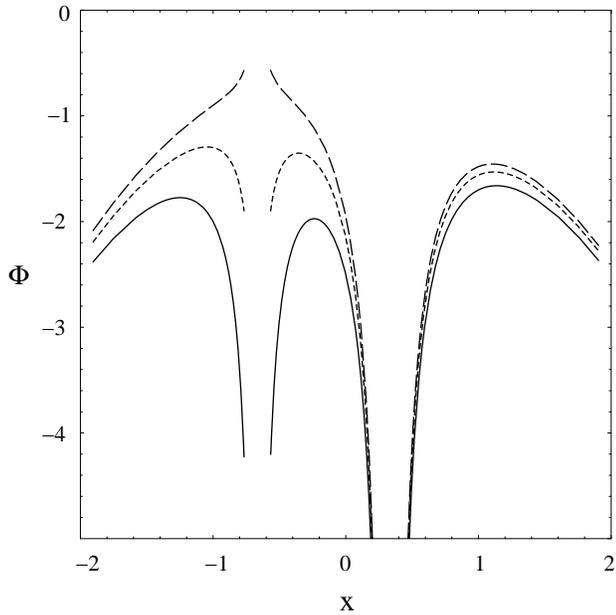}
  \caption{\label{Fig:global} Cross-sections through the effective
    equipotential surfaces (Eq.~\ref{Eq:phi_gen}) along the line joining
    the two stars, for $\mu = 1/3$. The mass-losing star (located at $x_1 =
    -2/3$) is the less massive component and has a (dimensionless) radius
    of 0.1. Three cases are depicted: $f = 0$ (solid line); $f = 0.7$
    (short-dashed line; note that the Roche lobe is becoming smaller as
    compared to the case $f = 0$); $f = 1.1$ (long-dashed line: there is no
    Roche lobe any longer).}
\end{figure}

Fig.~\ref{Fig:global} presents cross-cuts through the effective
equipotential surfaces (Eq.~\ref{Eq:phi_gen}) along the line joining the
two stars, for $\mu = 1/3$, i.e. the mass-losing star (located at $x_1 =
-2/3$) is the less massive component, and for three different values of
$f$: $f = 0$ (solid line), $f = 0.7$ (short-dashed line), and $f = 1.1$
(long-dashed line). The first case corresponds to a situation with no
extra-force, the second to a situation where the extra-force is present but
not sufficient to drive mass loss (note that the Roche lobe is becoming
smaller as compared to the case $f = 0$), and the third case corresponds to
a situation where the extra-force is actually driving the mass loss. Note
that in this case, there is no Roche lobe any longer.

\begin{figure*}
  \centering
  \begin{minipage}[c]{0.235\textwidth}
    \centering \includegraphics[width=\textwidth]{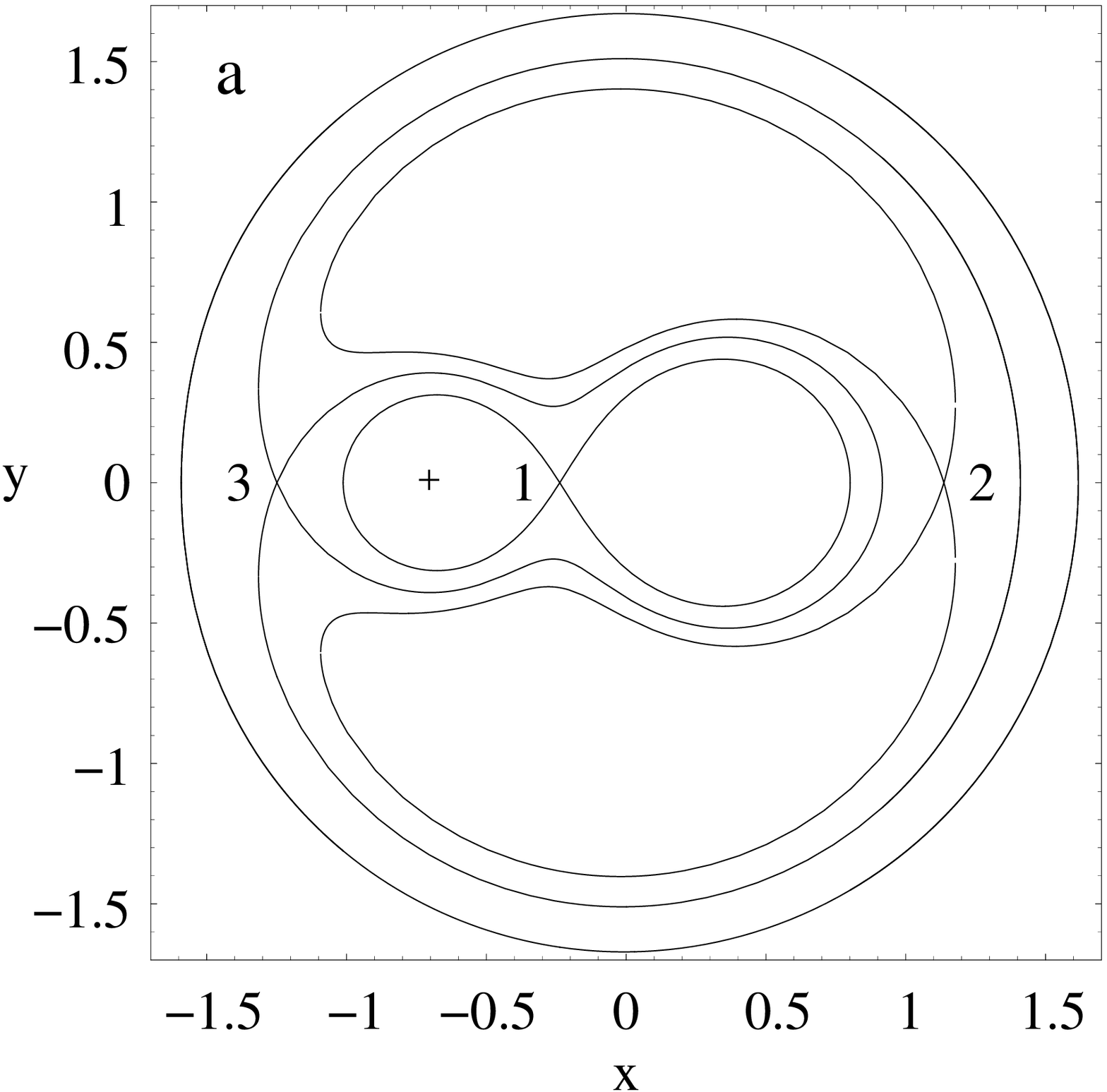}
  \end{minipage}
  \hspace{0.01\textwidth}
  \begin{minipage}[c]{0.235\textwidth}
    \centering \includegraphics[width=\textwidth]{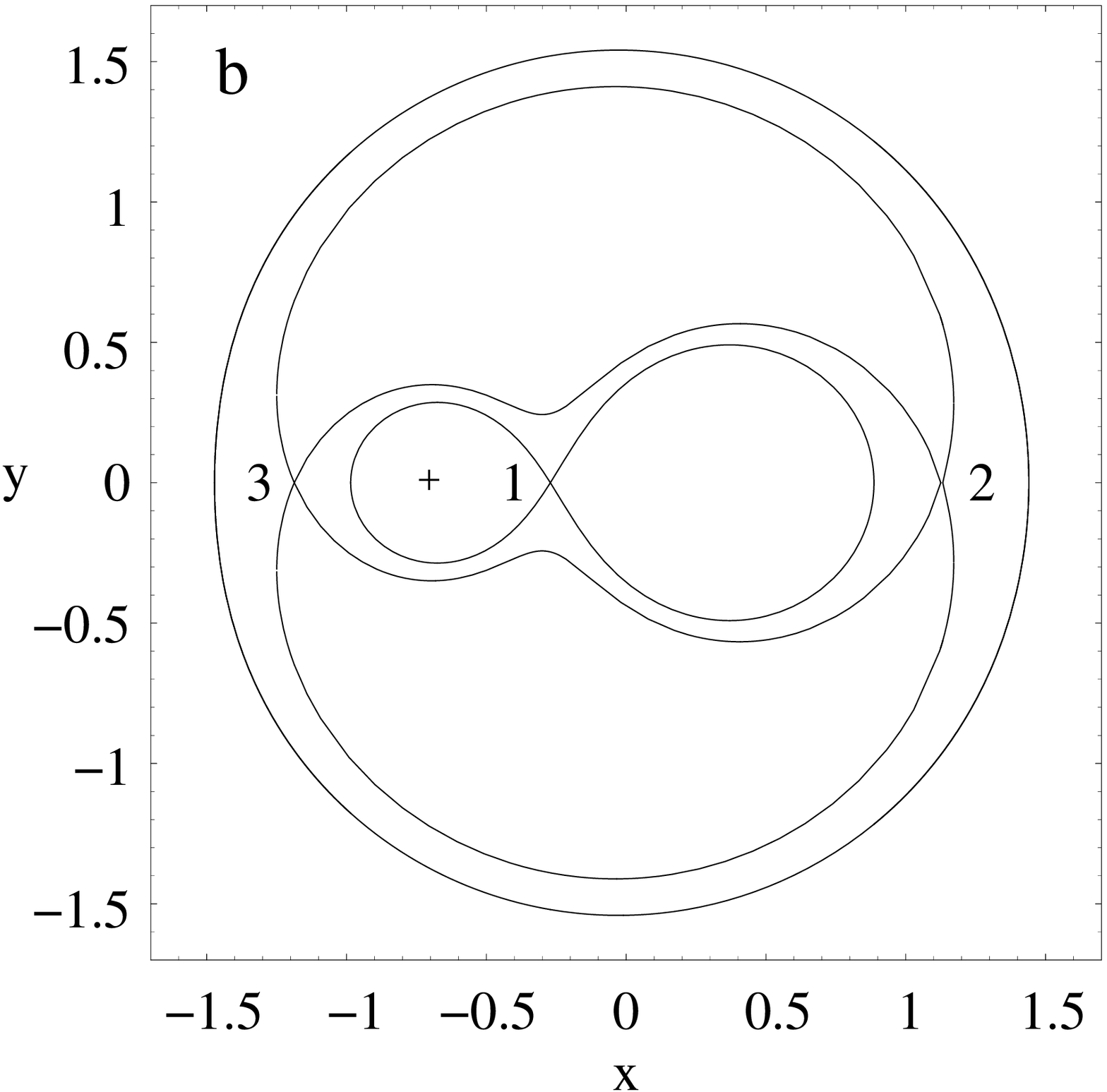}
  \end{minipage}
  \hspace{0.01\textwidth}
  \begin{minipage}[c]{0.235\textwidth}
    \centering \includegraphics[width=\textwidth]{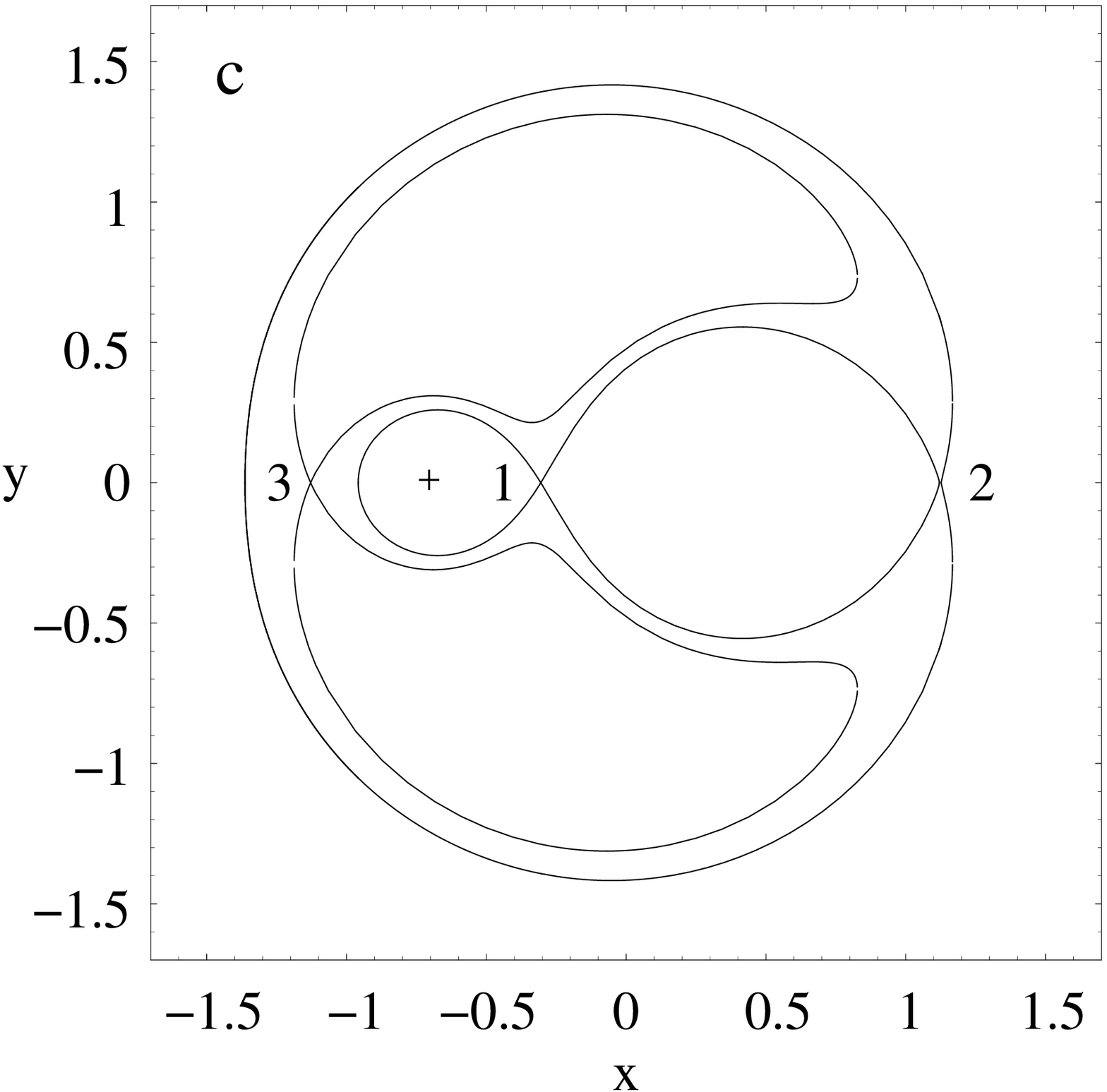}
  \end{minipage}
  \hspace{0.01\textwidth}
  \begin{minipage}[c]{0.235\textwidth}
    \centering \includegraphics[width=\textwidth]{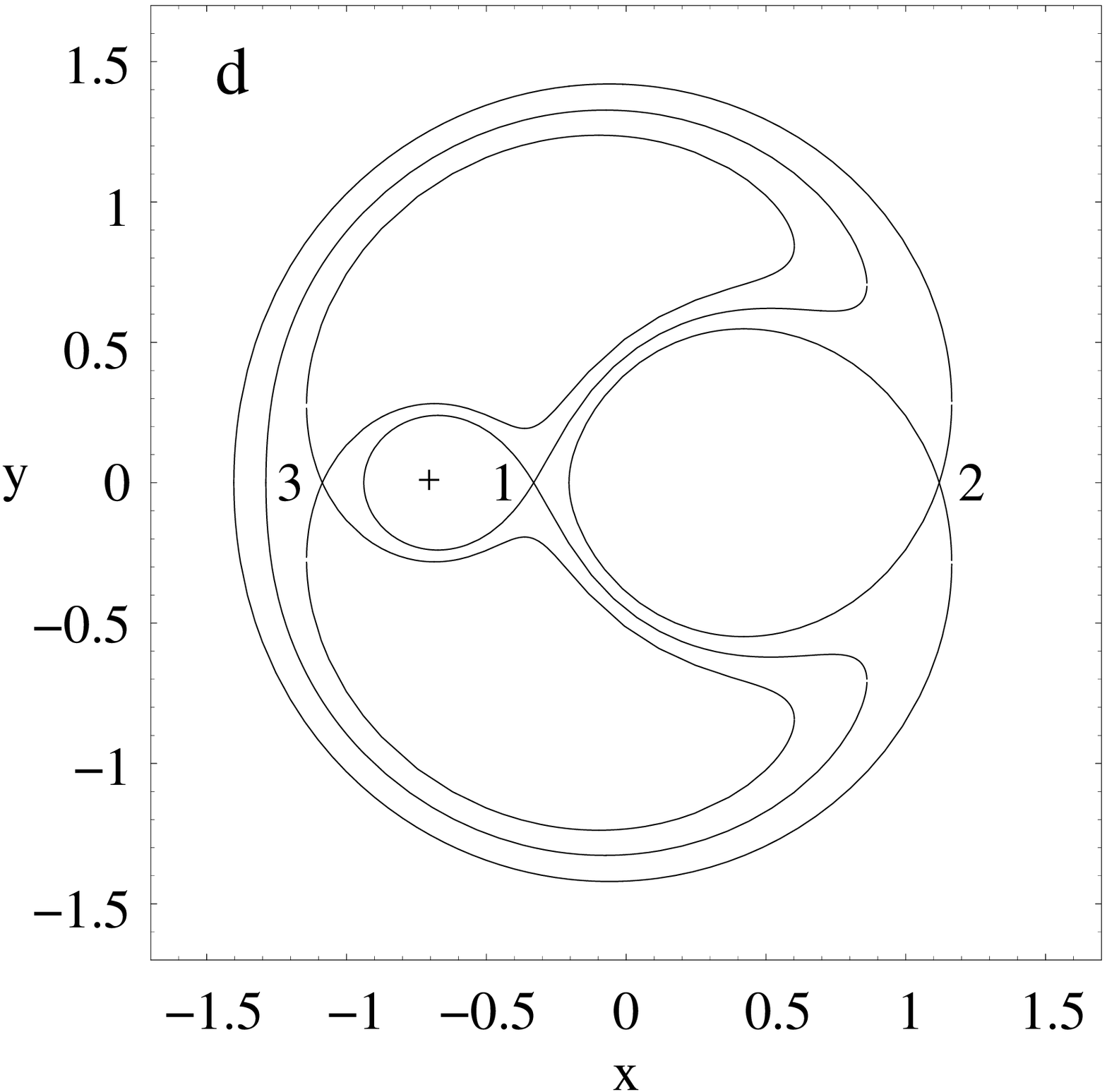}
  \end{minipage}
  
  \caption{\label{Fig:Equipotentials} Section in the orbital plane of the
    Roche equipotentials for a binary system with a dimensionless mass
    $\mu=1/3$ and $f=0$ (panel a), the critical values $f_1=0.27$ (panel
    b) and $f_2=0.48$ (panel c), obtained from numerical integration. The
    case $f=0.6$ is presented on panel d. The primary star, located at
    $x=\mu-1$, is depicted by the cross. The Lagrangian points \mbox{are
      labelled $1$, $2$ and $3$.}  }
  \label{Fig:equipotentials}  
\end{figure*}

Fig.~\ref{Fig:Equipotentials} presents the different families of
equipotentials as a function of $f$ and $\mu$, when $f < 1$.
Panels b and c correspond to the critical configurations where the
Lagrangian points $L_1$ and $L_3$ or $L_2$ and $L_3$ are located on the
same equipotentials.  These situations are encountered for the specific
values of $f$ denoted $f_1(\mu)$ and $f_2(\mu)$, respectively.  These
functions are approximated by the following expressions with a relative
error smaller than $1\%$ for $f_1(\mu)$ and $2\%$ (when $\mu\le 0.97$) for
$f_2(\mu)$:
\begin{equation}
  \label{Eq:Fit1}
  f_1(\mu)=\left(1-2 \mu\right) \left(1-0.80 \mu^{1/2}+0.82 \mu\right)
\end{equation}
\begin{equation}
  \label{Eq:Fit2}
  f_2(\mu)=\left(1-\mu\right) \left(1-0.72 \mu^{1/2}+0.66
  \mu^{3/2}+\mu^{17}\right)\textrm{\ .}
\end{equation}
The term $\mu^{17}$ is necessary to obtain a good fit to $f_2(\mu)$ when
$\mu$ approaches $1$.
\begin{figure}
  \centering
  \includegraphics[width=0.45\textwidth]{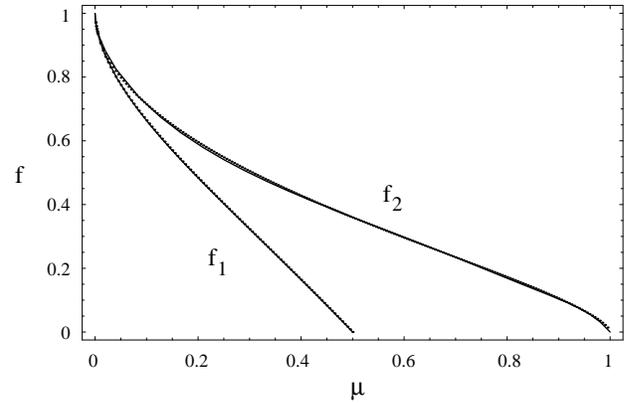}
  \caption{
    The functions $f_1(\mu)$ and $f_2(\mu)$ (see text), obtained by
    numerical integration (dots) and their analytical fit (solid curves) as
    expressed by Eqs.~\ref{Eq:Fit1} and \ref{Eq:Fit2}. The three 
    regions delineated by the curves in the $(f, \mu)$ plane correspond
    to three different topological structures for
    the equipotentials (see text and Fig.~\ref{Fig:Equipotentials}).
  }
  \label{Fig:fc}
\end{figure}

The different regions delineated by $f_1$ and $f_2$ in Fig.~\ref{Fig:fc}
correspond to different equipotential topologies.  Systems with $f<f_1$,
$f=f_1$, $f_1<f\le f_2$ and $f>f_2$ are topologically similar to cases
displayed in Fig.~\ref{Fig:Equipotentials} in panels a, b, c and d, respectively.
In the last case (panel d), the Roche
lobes of the two components do not belong to the same equipotential.
In contrast to the standard case, the matter ejected by the
primary is not necessarily transferred directly into the Roche lobe of the
companion, but all or a fraction of it may instead feed a circumbinary
disc. As mentioned in Sect.~\ref{Sect:Intro}, this
issue is important as such discs are very common in binaries
involving low- and intermediate-mass components.
The Coriolis force, which is not conservative,
plays also a major role in the formation of such discs, as shown by
numerical simulations
\citep{Theuns-Jorissen-93,Mastrodemos-1998,Mastrodemos-1999,Sytov-2009}.

\section{The RLOF criterion}
\label{Sect:RLOF}
\begin{figure}
  \centering
  \begin{minipage}[c]{0.45\textwidth}
    \centering \includegraphics[width=\textwidth]{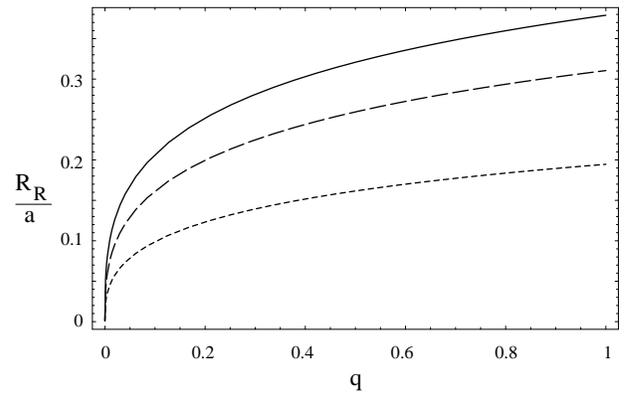}
  \end{minipage}
\centering
\caption{\label{Fig:Rr} The normalised Roche radius $R_R/a$ (where $a$ is
  the orbital separation) as a function of the mass ratio ($q\equiv
  M_1/M_2$) for the cases $f=0$ (solid line), $f=0.5$ (long-dashed line)
  and $f=0.9$ (short-dashed line).  }
\label{Fig:Rr}  
\end{figure}

\subsection{The modified Roche radius}
\label{Sect:Roche}
A generalisation to the cases $0\le f\le 1$ of the \citet{Eggleton-83}
formula  for the radius of a sphere with the same
volume as the Roche lobe, is given by
\begin{equation}
  \label{Eq.Rr}
        {\cal R}(q,f)\equiv\frac{R_R(q,f)}{a}=\frac{A(f)\ q^{2/3}}{B(f)\
          q^{2/3}+\ln\left(1+C(f)\ q^{1/3}\right)}
\end{equation}
where
\begin{displaymath}
  \begin{array}{ll}
    A(f)=&(1-f)^{1/3} (0.49+0.25 f+0.35 f^2\\
    &- 0.59 f^3+0.37 f^4) \textrm{,}\\
    B(f)=&0.6+0.3 f \textrm{,}\\
    C(f)=&1+f \textrm{,}
  \end{array}
\end{displaymath}
$q\equiv M_1/M_2$ is the mass ratio of the two components (star 1 being the
one with non-negligible radiation pressure) and $a$ is the orbital
separation. For $f=0$, the \mbox{\cite{Eggleton-83}} expression is
recovered.
Fig.~\ref{Fig:Rr} shows the Roche radius as a function of the
mass ratio $q$ for the values $f=0$, $f=0.5$ and $f=0.9$. The Roche radius
is clearly smaller in the presence of radiation pressure and is reduced by
a factor of $\sim2$ between the cases $f=0$ and $f=0.9$.
The Roche lobe vanishes (and its radius thus goes to
zero) when the radiation force becomes equal to the gravitational
attraction (i.e. $f=1$).

The analytical fit to the Roche radius as approximated by Eq.~\ref{Eq.Rr}
is accurate to better than $3 \%$ over the extended range $0.1 \le q \le
\infty$ and $0 \le f \le 0.95$.
Outside this parameter range, the relative error is less than $7 \%$.  The
fit is based on numerical results obtained with the method outlined by
\citet{Huang-1990}.
We emphasize once again that, according to the
discussion of Sect.~\ref{Sect:radiation}, the Roche radius expressed by
Eq.~\ref{Eq.Rr} applies only to stars with an extended atmosphere
where the free-streaming approximation holds for the radiation field.

\subsection{RLOF stability}
\label{Sect:RLOF stability}
We now evaluate the impact of the above modifications on the RLOF
stability. In the Roche model, the stability condition imposes that, when
the star fills its Roche lobe, subsequent mass loss does not lead to a
runaway situation. It is expressed by the condition \mbox{$\zeta_R<\zeta$},
where
\begin{equation}
  \zeta_R\equiv\frac{\ud\ln R_R}{\ud\ln M_1}
\end{equation}
and
\begin{equation}
  \zeta\equiv\frac{\ud\ln R_1}{\ud\ln M_1}
\end{equation}
are the Roche-lobe mass-radius exponent and the mass-radius exponent of the
donor, respectively. A star responds to mass loss on two timescales. The
immediate response is on the adiabatic time scale ($\tau_\mathrm{dyn}$),
after which hydrostatic equilibrium is restored but negligible heat
transport has occurred. The mass-radius exponent characterizing this
adiabatic readjustment is denoted by $\zeta_\mathrm{ad}$. On the other hand,
thermal equilibrium is recovered on the Kelvin-Helmoltz time scale
($\tau_\mathrm{KH}$) and is characterised by the exponent
$\zeta_\mathrm{th}$.  The adiabatic and thermal stability conditions
become respectively
\begin{equation}
  \zeta_R<\zeta_\mathrm{ad}\textrm\ {}
\end{equation}
and
\begin{equation}
  \zeta_R<\zeta_\mathrm{th}\textrm\ {.}
\end{equation}
If neither of these conditions is satisfied, mass transfer proceeds on the fastest of the two timescales.

To estimate $\zeta_R$, \citet{Soberman-1997} \citep[see
  also][]{Jorissen-2003} assume that a fraction $\alpha$ of the mass lost in
the wind escapes to infinity, that a fraction $\eta$ is accreted by the
companion and the remaining $\delta = 1-\alpha-\eta$ goes into feeding a
circumbinary disc of radius $a_r=\gamma^2 a$. In this framework the
Roche-lobe mass-radius exponent writes
\begin{displaymath}
  \zeta_R=2\left(\frac{\alpha}{1+q} +\gamma\delta(1+q)
  -\lambda\right)+\frac{q}{1+q}(1-\eta)
\end{displaymath}
\begin{equation}
  \label{Eq:zeta}
  \hspace{0.8cm} +2 (\eta q-1)+\frac{\ud\ln {\cal R}}{\ud\ln q} (1+\eta
  q)\textrm\ {,}
\end{equation}
where $\lambda\sim 0.05$ according to the hydrodynamical simulations of
\cite{Theuns-1996}.

When an extra force is present, the term $(\ud\ln {\cal R}/\ud\ln q)$ increases
and may possibly destabilise the system. However, the effect is small ($< 10 \%$
for $f$ ranging between $0$ and $0.9$) and the stability condition remains
always dominated by the first three terms of Eq.~\ref{Eq:zeta},
which depend on the efficiency of mass transfer through the parameters
$\eta, \delta$ and $\gamma$. This is apparent from Fig.~4 of
\cite{Soberman-1997} or Fig.~9.14 of \cite{Jorissen-2003}.  The
direct effect of the
extra force on RLOF stability is consequently negligible.

\subsection{Critical period} 
\label{Sect:Critical period}

An interesting application of Eq.~\ref{Eq.Rr} is related to the
\emph{critical period} $P_\mathrm{crit}$, the orbital period below which
RLOF occurs, given by \cite[]{Eggleton-2006}

\begin{displaymath}
  P_{\mathrm{crit}}= \left( \frac{4\pi^2 R^3}{G M}
  \right)^{1/2} {\cal R}(q,f)^{-3/2}
\end{displaymath}
\begin{equation}\vspace{-0.5cm}
  \label{Eq.Pcrit}
  \hspace{0.8cm}\sim 0.1159 \left(
  \frac{R^3(R_\odot)}{M(M_\odot)} \right)^{1/2} {\cal R}(q,f)^{-3/2}
  \textrm{\ (days) ,}
\end{equation}
where $M=M_1+M_2$.  $P_{\mathrm{crit}}$ increases strongly when an extra
force is present as shown in Fig.~\ref{Fig:Pcrit}.
This means that RLOF, either stable or not, would occur at a longer
orbital period than inferred from the size of the classical Roche lobe.
Such an effect of the extra force (with an estimated $f=0.65$ -- $0.85$)
shrinking the effective Roche lobe was offered by \citet{Frankowski-2001}
as an explanation for the symbiotic stars ``avoiding'' to fill their Roche
lobes \citep{Muerset-1999}. As noted in Sect.~\ref{Sect:radiation}, it was
also discovered that a number of symbiotics exhibit ellipsoidal variations,
apparently being tidally distorted despite not quite filling their classical
Roche lobes \citep{Mikolajewska-2007}.
Observations of the orbital circularisation in a sample of binaries with M
giant primaries supply another possible example of this effect: these giants
do not fill more than $\sim 0.5$ of their classical Roche lobe
\citep{Frankowski-2009}. If this interpretation of these systems
properties being modified by the extra force is correct, the $f>0.7$ values implied for
these cool giants are significantly higher than the estimates given for
absorption/scattering in molecular lines in the photospheres
of AGB stars (see Sect.~\ref{Sect:Determination of f values}).
They are closer to the effective $f \sim 1$ values inferred from the
pulsation-driven winds of AGB Mira stars.
This reflects the fact that radiation pressure is not the only
energy and momentum source available in the photospheres of giant stars
\cite[e.g., even non-pulsating cool giants exhibit
bulk motions, manifesting as radial-velocity jitter;][]
{Gunn-Griffin-1979}.

\begin{figure}
  \centering
  \begin{minipage}[c]{0.45\textwidth}
    \centering \includegraphics[width=\textwidth]{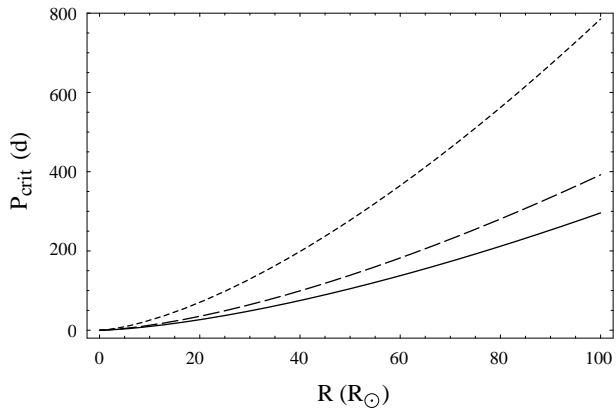}
  \end{minipage}
  \centering
  
  \caption{\label{Fig:Pcrit} Critical period $P_\mathrm{crit}$ in days from
    Eq.~\ref{Eq.Pcrit} as a function of the stellar radius $R (R_\odot)$ for
    $f=0$ (solid line), $f=0.5$ (long-dashed line) and $f=0.9$
    (short-dashed line). In each case, $M_1=1.2\; M_\odot$ and $M_2=0.6\;
    M_\odot$ (i.e., $q=2$ or $\mu=2/3$).
  }
  \label{Fig:Pcrit}  
\end{figure}

\subsection{No RLOF for $\bm{f > 1}$}
\label{Sect:Mass losing stars}
We showed in Fig.~\ref{Fig:global} that when $f > 1$, there is no longer
a critical Roche surface around the mass-losing star, so that the very concept 
of RLOF becomes meaningless.

For a star where, e.g., the radiation pressure is large enough to expel stellar
material, the potential corresponds to a net repulsive force (long-dashed-line in
Fig.~\ref{Fig:global}) and {\em} the Roche lobe around the mass-losing star
has no meaning any longer!  Similarly the stellar radius needs to be
re-defined, especially in the case of optically-thick winds (like for WR
stars), when the photospheric radius (corresponding to an optical depth
$\tau = 2/3$) falls within the wind \citep{deLoore-1982,Baschek-1991,Moffat-1996}.
This property is accounted for in recent stellar models which correct the stellar
radius using extrapolation of the wind expansion law in the optically-thick region
\citep[]{Langer-1989,Hamann-1993}.

A further consequence of the absence of a Roche lobe around radiatively
driven mass-losing stars is that there is no dramatic change in the
mass-loss regime from wind mass loss to RLOF, as the latter is now
ill-defined.
In fact, several authors have already promoted this idea of a smooth
transition of the mass loss rate from the wind to the RLOF regime
\citep[]{Tout-Eggleton-88,Frankowski-2001}.

An important issue related to mass transfer is to evaluate whether or not a
common envelope will form and this outcome depends on the geometry of the
equipotentials.  Contrarily to the situation prevailing during classical
RLOF, the mass lost by the wind is not necessarily injected into the Roche
lobe of the companion as illustrated in Fig.~\ref{Fig:Equipotentials of
  mass losing star}. In particular, a substantial fraction of the wind can
avoid the companion's Roche lobe and instead be used to form a circumbinary
disc.
\begin{figure}
  \hspace{.125\textwidth}
  \includegraphics[width=0.375\textwidth]{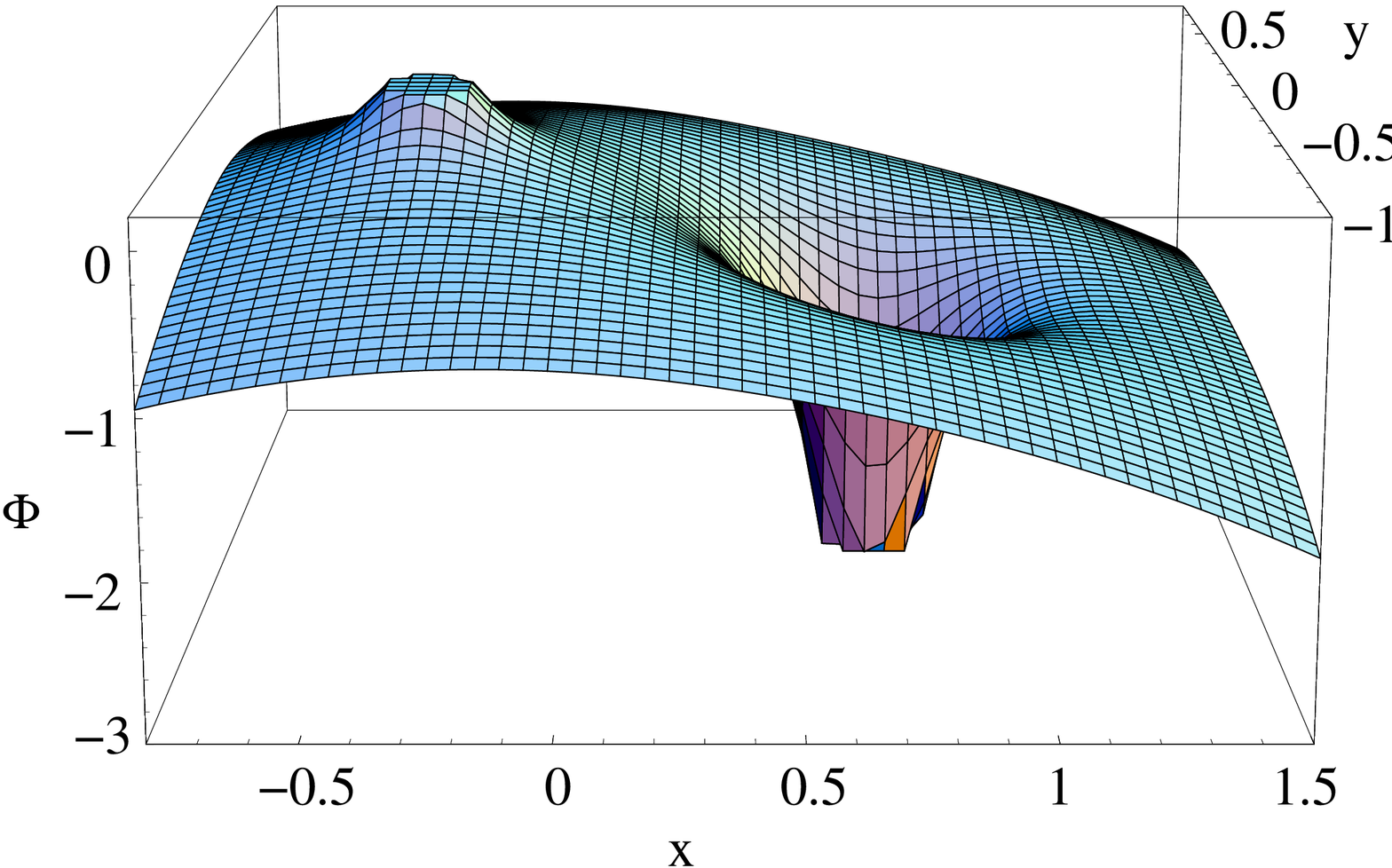}
  \includegraphics[width=0.46\textwidth]{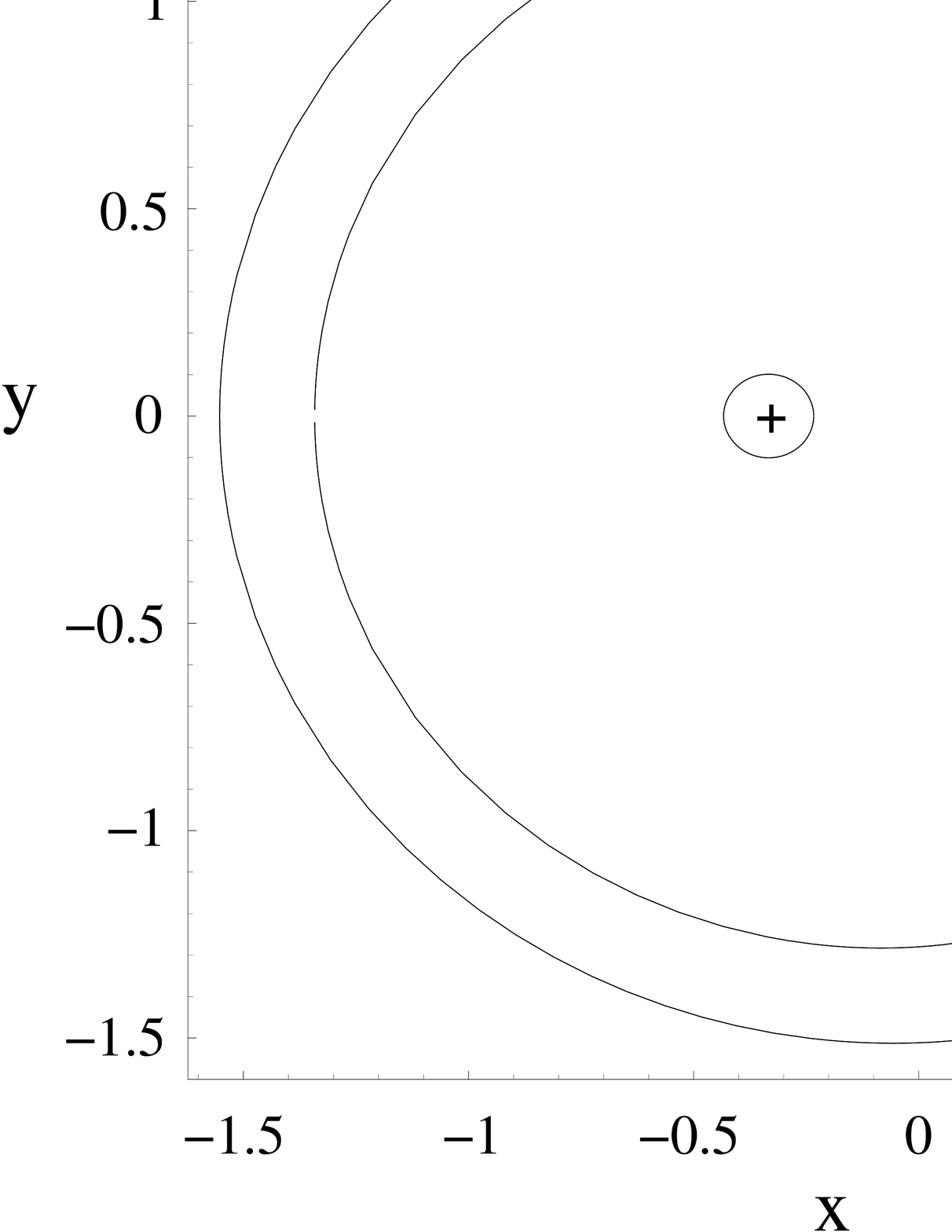}
  \caption{\label{Fig:Equipotentials of mass losing star} Roche potential
    surface (upper panel) and equipotentials on the orbital plane (bottom
    panel) for $\mu =2/3$ and \mbox{$f=1.1$}. The mass-losing star, located
    at $x=\mu-1$ and depicted by the cross, has a radius of $0.1$ (in units
    of the orbital separation). The only Lagrangian point is $L_2$.  }
\end{figure}

\subsection{Pulsation-driven winds: The case of Mira stars}
\label{Sect:Pulsation driven}
\begin{figure}
\centering
\includegraphics[width=0.45\textwidth]{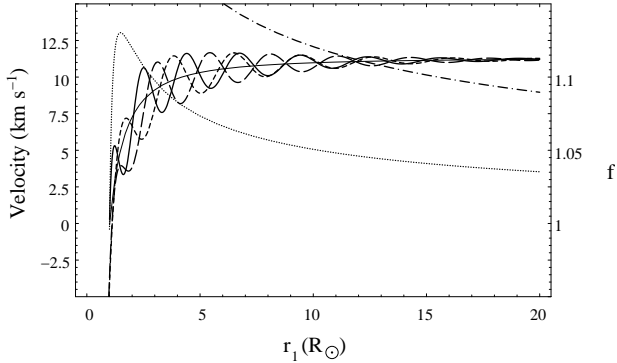}
\caption{\label{Fig:wind} Wind velocities $v_{\rm wind}(r_1)$ (solid,
  short- and long-dashed lines and left-hand scale) and escape velocity
  (dot-dashed line) from model calculations of \cite{Bowen-88} for a Mira
  star of $M=1.2\ M_\odot$, $R=270\ R_\odot$ and $L=5415\ L_\odot$,
  including pulsation and radiation pressure on dust. The function $f(r_1)$
  (dotted line and right-hand scale) is derived from Eq.~\ref{Eq:f}.  }
\label{Fig:wind}
\end{figure}
The process driving the wind in Mira stars seems to start with
pulsation-induced shock waves that lift the matter high enough above the
photosphere for dust to form. Since all photospheric particles will feel
this upward force, it should be included in the effective potential. 
However, in the case of momentum
transfer from shock waves, there is no simple mathematical expression for
the extra acceleration $g_{\rm ext}$, unlike the case of the
radiation-pressure force. Nevertheless, it is possible to infer the run of
$g_{\rm ext}$ as a function of the distance $r_1$ from the stellar surface
using model predictions for the wind velocity \citep[see
  e.g.,][]{Bowen-88,Willson-00}. 
In the steady state approximation, the acceleration d$v$/d$t$ for a wind
particle writes
\begin{equation}
  -g_{\rm grav} + g_{\rm ext} = {\mathrm d}v/{\mathrm d}t = v\; \frac{{\mathrm d}v}{{\mathrm d}r_1}\\
\end{equation}
The above relation implicitly assumes that the wind velocity function
is derivable. It thus requires to smooth the discontinuities associated
with the shock waves (thick line in Fig.~\ref{Fig:wind}). By suppressing the
non-conservative character of the shock waves, this case becomes similar
to that of a conservative force deriving from a potential $\Phi_{\rm
  ext}$.
The fact that Mira stars have relatively low-velocity winds
implies that this extra force is mostly used to lift the matter out of the
potential well.
The run of $f$ with distance then writes 
\begin{equation}
  \label{Eq:f}
  f(r_1) \equiv \frac{g_{\rm ext}(r_1)}{g_{\rm grav}(r_1)}= 1 + 
\frac{v}{g_{\rm grav}(r_1)}\;\frac{{\mathrm d}v}{{\mathrm d}r_1} 
  \mathrm{.}
\end{equation}

Because Mira stars are pulsating, their wind-velocity curves are
time-dependent \citep[][]{Bowen-88}.  In order to mimic this time
variability, high-frequency spatial and temporal variations are added to
the smooth, long-range velocity curve. The high-frequency component
corresponds to a sinusoidal curve whose amplitude decreases exponentially
with distance from the stellar surface. This high-frequency component is
moreover phase-shifted by $\pi/2$ (solid line in Fig.~\ref{Fig:wind}) and
$\pi$ (long-dashed line) to mimick the temporal evolution (shock-wave propagation).
The corresponding potentials are represented in Fig.~\ref{Fig:Mira
  potential} (solid, long- and short-dashed lines).  Inside the stellar
radius, matter is supposed to be in hydrostatic equilibrium, with gas and
radiation pressure balancing gravitation, so that matter is at rest on
average. 
  At different times of the shock-wave propagation, the
potential can be repulsive (solid and long-dashed curves), allowing for the
ejection of matter, or becomes attractive (short-dashed line), temporarily
preventing mass ejection. 

In our calculations, $f$ remains on average only
slightly larger than unity which
confirms the fact that the driving mechanism is mainly used to work against
the gravitational attraction.
Finally, for binary systems involving Mira stars, the companion does
influence the mass-losing star by altering the
physical processes driving its wind. \citet{Frankowski-2001} for instance
stress that the empirical formulae fitting red-giant mass-loss rates
\citep{Reimers-1975,Arndt-1997} depend on the surface gravity, and
consider how gravity of the mass-losing star will be altered, both because of the
straightforward addition of the companion gravitational attraction and
because of the tidal distortion. Furthermore, in Mira stars, the
gravitational field of the companion may also alter the source of the
mass-loss, by disturbing the formation and the properties of the shock
waves. All these aspects contribute to making the
determination of $f$ rather uncertain.

\begin{figure}
  \centering
  \includegraphics[width=0.45\textwidth]{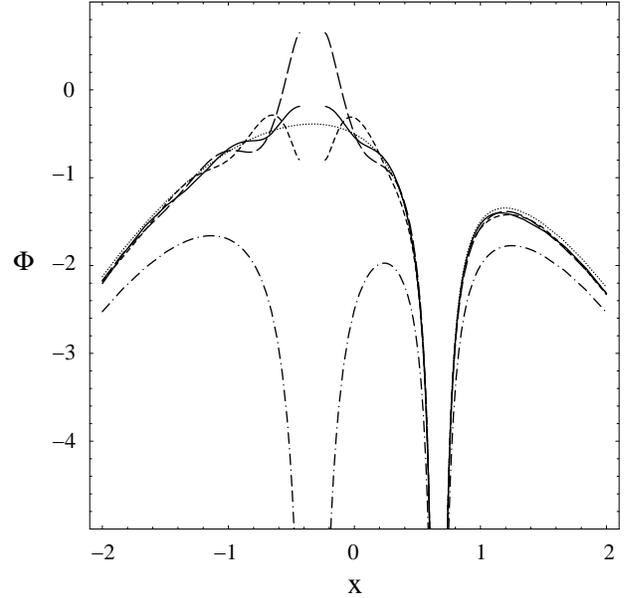}
  \caption{\label{Fig:Mira potential} Same as Fig.~\ref{Fig:global} in the
    case of a Mira-type wind (solid, short- and long-dashed lines), derived
    from the wind velocity curves displayed on Fig.~\ref{Fig:wind} and
    using Eq.~\ref{Eq:f}. The star is assumed to have a radius of 0.1 (in
    units of the orbital separation), and $\mu=2/3$. 
    The dotted and dot-dashed lines correspond respectively to
    the cases $f=1$ and $f=0$.  }
\end{figure}

\section{Conclusions}
\label{Sect:Conclusions}

This paper presents and analyses the Roche potential modified by the
presence of an extra force associated with radiation pressure or
pulsation. The magnitude of this perturbing force is quantified by the
parameter $f$ which represents the ratio of the extra-force and the
gravitational attraction \citep{Schuerman-1972}. An estimate of this
parameter for main sequence, RGB, AGB and Mira stars is also provided.

For $0< f < 1$, the Roche potential may be substantially modified. In
particular, if $f > f_1$ the deformation of the equipotentials allow the
matter ejected by the mass-losing star to go into a circumbinary disc.  As
the extra force ($f$) becomes stronger, the Roche radius decreases,
favouring RLOF mass transfer. 

Numerical fits and generalisation of the Roche radius are provided in this
paper for $f<1$. It is shown that the effects of the extra force on the
RLOF stability is negligible.

For $f>1$, the Roche lobe has no meaning any longer. Such situations
occur in luminous stars where radiation drives the mass loss or in
pulsating giant stars. In this latter case, the recurrent deposition of
momentum by the shock waves in the atmosphere allows matter at the surface
to be expelled.

The consideration of a modified Roche lobe is (directly or indirectly) 
supported by numerous
  observations (like the frequent occurrence of circumbinary discs in post-mass-transfer systems, 
  and the small classical Roche-filling factors derived for symbiotic or M giants despite their 
  ellipsoidal variability or circular orbit...) and should be taken into account.

\acknowledgements{
  L.S. is Research Associate from FRS-F.N.R.S., and T.D. is 
  {\it Boursier F.R.I.A.} This work has been partly funded by an {\it Action de recherche concert\'ee (ARC)} from the {\it Direction g\'en\'erale de l'Enseignement non obligatoire et de la Recherche scientifique -- Direction de la recherche scientifique -- Communaut\'e fran\c caise de Belgique.}
  

\begin{thebibliography}{73}
\expandafter\ifx\csname natexlab\endcsname\relax\def\natexlab#1{#1}\fi

\bibitem[{{Abbott}(1982)}]{Abbott-1982}
{Abbott}, D.~C. 1982, \apj, 259, 282

\bibitem[{{Alecian} \& {LeBlanc}(2002)}]{Alecian-2002}
{Alecian}, G. \& {LeBlanc}, F. 2002, \mnras, 332, 891

\bibitem[{{Anderson} \& {Shu}(1977)}]{Anderson-1977}
{Anderson}, L. \& {Shu}, F.~H. 1977, \apj, 214, 798

\bibitem[{{Arndt} {et~al.}(1997){Arndt}, {Fleischer}, \&
  {Sedlmayr}}]{Arndt-1997}
{Arndt}, T.~U., {Fleischer}, A.~J., \& {Sedlmayr}, E. 1997, \aap, 327, 614

\bibitem[{{Babel}(1992)}]{Babel-1992}
{Babel}, J. 1992, \aap, 258, 449

\bibitem[{{Baschek} {et~al.}(1991){Baschek}, {Scholz}, \&
  {Wehrse}}]{Baschek-1991}
{Baschek}, B., {Scholz}, M., \& {Wehrse}, R. 1991, \aap, 246, 374

\bibitem[{{Bowen}(1988)}]{Bowen-88}
{Bowen}, G.~H. 1988, \apj, 329, 299

\bibitem[{{Castor} {et~al.}(1975){Castor}, {Abbott}, \& {Klein}}]{Castor-75}
{Castor}, J.~I., {Abbott}, D.~C., \& {Klein}, R.~I. 1975, \aj, 195, 157

\bibitem[{{Cox}(2000)}]{Cox-2000}
{Cox}, A.~N. 2000, Allen's Astrophysical Quantites (Athlone Press)

\bibitem[{{de Loore} {et~al.}(1982){de Loore}, {Hellings}, \&
  {Lamers}}]{deLoore-1982}
{de Loore}, C., {Hellings}, P., \& {Lamers}, H.~J.~G. 1982, in IAU Symposium,
  Vol.~99, Wolf-Rayet Stars: Observations, Physics, Evolution, ed. C.~W.~H. {de
  Loore} \& A.~J. {Willis}, 53--56

\bibitem[{{de Ruyter} {et~al.}(2006){de Ruyter}, {van Winckel}, {Maas}, {Lloyd
  Evans}, {Waters}, \& {Dejonghe}}]{DeRuyter-2006}
{de Ruyter}, S., {van Winckel}, H., {Maas}, T., {et~al.} 2006, \aap, 448, 641

\bibitem[{{Djurasevic}(1986)}]{Djurasevic-1986}
{Djurasevic}, G. 1986, \apss, 124, 5

\bibitem[{{Drechsel} {et~al.}(1995){Drechsel}, {Haas}, {Lorenz}, \&
  {Gayler}}]{Drechsel-1995}
{Drechsel}, H., {Haas}, S., {Lorenz}, R., \& {Gayler}, S. 1995, \aap, 294, 723

\bibitem[{{Eggleton}(2006)}]{Eggleton-2006}
{Eggleton}, P. 2006, Evolutionary Processes in Binary and Multiple Stars
  (Cambridge University Press)

\bibitem[{{Eggleton}(1983)}]{Eggleton-83}
{Eggleton}, P.~P. 1983, \apj, 268, 368

\bibitem[{{Elitzur} {et~al.}(1989){Elitzur}, {Brown}, \&
  {Johnson}}]{Elitzur-1989}
{Elitzur}, M., {Brown}, J.~A., \& {Johnson}, H.~R. 1989, \apjl, 341, L95

\bibitem[{{Frankowski}(2009)}]{Frankowski-2009a}
{Frankowski}, A. 2009, in Asymmetrical planetary nebulae IV, ed. R.~L.~M.
  {Corradi}, A.~{Manchado}, \& N.~{Soker} (I.A.C. electronic publication),
  501--508

\bibitem[{{Frankowski} {et~al.}(2009){Frankowski}, {Famaey}, {van Eck},
  {Mayor}, {Udry}, \& {Jorissen}}]{Frankowski-2009}
{Frankowski}, A., {Famaey}, B., {van Eck}, S., {et~al.} 2009, \aap, 498, 479

\bibitem[{{Frankowski} \& {Jorissen}(2007)}]{Frankowski-2007a}
{Frankowski}, A. \& {Jorissen}, A. 2007, Baltic Astronomy, 16, 104

\bibitem[{{Frankowski} \& {Tylenda}(2001)}]{Frankowski-2001}
{Frankowski}, A. \& {Tylenda}, R. 2001, A\&A, 367, 513

\bibitem[{{Friend} \& {Castor}(1982)}]{Friend-1982}
{Friend}, D.~B. \& {Castor}, J.~I. 1982, \apj, 261, 293

\bibitem[{{Gail} \& {Sedlmayr}(1987)}]{Gail-87a}
{Gail}, H.~P. \& {Sedlmayr}, E. 1987, \aap, 171, 197

\bibitem[{{Gayley}(1995)}]{Gayley-1995}
{Gayley}, K.~G. 1995, \apj, 454, 410

\bibitem[{{Glatzel} {et~al.}(1993){Glatzel}, {Kiriakidis}, \&
  {Fricke}}]{Glatzel-93}
{Glatzel}, W., {Kiriakidis}, M., \& {Fricke}, K.~J. 1993, \mnras, 262, L7

\bibitem[{{Gunn} \& {Griffin}(1979)}]{Gunn-Griffin-1979}
{Gunn}, J.~E. \& {Griffin}, R.~F. 1979, \aj, 84, 752

\bibitem[{{Gustafsson} {et~al.}(2008){Gustafsson}, {Edvardsson}, {Eriksson},
  {J{\o}rgensen}, {Nordlund}, \& {Plez}}]{Gustafsson-2008}
{Gustafsson}, B., {Edvardsson}, B., {Eriksson}, K., {et~al.} 2008, \aap, 486,
  951

\bibitem[{{Hamann}(1993)}]{Hamann-1993}
{Hamann}, W.-R. 1993, Space Science Reviews, 66, 237

\bibitem[{{Holzer} \& {MacGregor}(1985)}]{Holzer-1985}
{Holzer}, T.~E. \& {MacGregor}, K.~B. 1985, in Astrophysics and Space Science
  Library, Vol. 117, Mass Loss from Red Giants, ed. M.~{Morris} \&
  B.~{Zuckerman}, 229--255

\bibitem[{{Howarth}(1997)}]{Howarth-1997}
{Howarth}, I.~D. 1997, The Observatory, 117, 335

\bibitem[{{Howarth} \& {Smith}(2001)}]{Howarth-2001}
{Howarth}, I.~D. \& {Smith}, K.~C. 2001, \mnras, 327, 353

\bibitem[{{Huang} \& {Taam}(1990)}]{Huang-1990}
{Huang}, R.~Q. \& {Taam}, R.~E. 1990, \aap, 236, 107

\bibitem[{{Hui-Bon-Hoa} {et~al.}(2001){Hui-Bon-Hoa}, {LeBlanc}, {Hauschildt},
  \& {Baron}}]{HuiBonHoa-2002}
{Hui-Bon-Hoa}, A., {LeBlanc}, F., {Hauschildt}, P., \& {Baron}, E. 2001, \aap,
  377, 175

\bibitem[{{Jorgensen} \& {Johnson}(1992)}]{Jorgensen-1992}
{Jorgensen}, U.~G. \& {Johnson}, H.~R. 1992, \aap, 265, 168

\bibitem[{{Jorissen}(2003)}]{Jorissen-2003}
{Jorissen}, A. 2003, in Asymptotic Giant Branch Stars, ed. H.~{Habing} \&
  H.~{Olofsson} (New York: Springer Verlag), 461--518

\bibitem[{{Kondo} \& {McCluskey}(1976)}]{Kondo-1976}
{Kondo}, Y. \& {McCluskey}, G.~E. 1976, in IAU Symposium, Vol.~73, Structure
  and Evolution of Close Binary Systems, ed. P.~{Eggleton}, S.~{Mitton}, \&
  J.~{Whelan}, 277--282

\bibitem[{{Kudritzki} \& {Puls}(2000)}]{Kudritzki-00}
{Kudritzki}, R.-P. \& {Puls}, J. 2000, \araa, 38, 613

\bibitem[{{Lamers}(1997)}]{Lamers-1997}
{Lamers}, H.~J.~G.~L.~M. 1997, in Lecture Notes in Physics, Berlin Springer
  Verlag, Vol. 497, Stellar Atmospheres: Theory and Observations, ed. J.~P. {de
  Greve}, R.~{Blomme}, \& H.~{Hensberge}, 69

\bibitem[{{Langer}(1989)}]{Langer-1989}
{Langer}, N. 1989, \aap, 210, 93

\bibitem[{{Lemke}(1990)}]{Lemke-1990}
{Lemke}, M. 1990, \aap, 240, 331

\bibitem[{{Maeder} \& {Meynet}(2000)}]{Maeder-Meynet-2000}
{Maeder}, A. \& {Meynet}, G. 2000, \aap, 361, 159

\bibitem[{{Mastrodemos} \& {Morris}(1998)}]{Mastrodemos-1998}
{Mastrodemos}, N. \& {Morris}, M. 1998, ApJ, 497, 303

\bibitem[{{Mastrodemos} \& {Morris}(1999)}]{Mastrodemos-1999}
{Mastrodemos}, N. \& {Morris}, M. 1999, ApJ, 523, 357

\bibitem[{{Michaud} \& {Charland}(1986)}]{Michaud-1986}
{Michaud}, G. \& {Charland}, Y. 1986, \apj, 311, 326

\bibitem[{{Michaud} {et~al.}(1983){Michaud}, {Tarasick}, {Charland}, \&
  {Pelletier}}]{Michaud-1983}
{Michaud}, G., {Tarasick}, D., {Charland}, Y., \& {Pelletier}, C. 1983, \apj,
  269, 239

\bibitem[{{Miko\l ajewska}(2007)}]{Mikolajewska-2007}
{Miko\l ajewska}, J. 2007, Baltic Astron., 16, 1

\bibitem[{{Moffat} \& {Marchenko}(1996)}]{Moffat-1996}
{Moffat}, A.~F.~J. \& {Marchenko}, S.~V. 1996, \aap, 305, L29+

\bibitem[{{M{\"u}rset} \& {Schmid}(1999)}]{Muerset-1999}
{M{\"u}rset}, U. \& {Schmid}, H.~M. 1999, A\&AS, 137, 473

\bibitem[{{Nugis} \& {Lamers}(2000)}]{Nugis-2000}
{Nugis}, T. \& {Lamers}, H.~J.~G.~L.~M. 2000, \aap, 360, 227

\bibitem[{{Owocki}(2004)}]{Owocki-2004}
{Owocki}, S. 2004, in EAS Publications Series, Vol.~13, Evolution of Massive
  Stars, Mass Loss and Winds, ed. M.~{Heydari-Malayeri}, P.~{Stee}, \& J.-P.
  {Zahn}, 163--250

\bibitem[{{Owocki}(2007)}]{Owocki-2007}
{Owocki}, S. 2007, in Astronomical Society of the Pacific Conference Series,
  Vol. 367, Massive Stars in Interactive Binaries, ed. N.~{St.-Louis} \&
  A.~F.~J. {Moffat}, 233

\bibitem[{{Owocki} \& {Gayley}(1999)}]{Owocki-99}
{Owocki}, S.~P. \& {Gayley}, K.~G. 1999, in IAU Symposium, Vol. 193, Wolf-Rayet
  Phenomena in Massive Stars and Starburst Galaxies, ed. K.~A. {van der Hucht},
  G.~{Koenigsberger}, \& P.~R.~J. {Eenens}, 157

\bibitem[{{Pauldrach} {et~al.}(1986){Pauldrach}, {Puls}, \&
  {Kudritzki}}]{Pauldrach-1986}
{Pauldrach}, A., {Puls}, J., \& {Kudritzki}, R.~P. 1986, \aap, 164, 86

\bibitem[{{Phillips} \& {Podsiadlowski}(2002)}]{Phillips-2002}
{Phillips}, S.~N. \& {Podsiadlowski}, P. 2002, \mnras, 337, 431

\bibitem[{{Puls} {et~al.}(2000){Puls}, {Springmann}, \& {Lennon}}]{Puls-2000}
{Puls}, J., {Springmann}, U., \& {Lennon}, M. 2000, \aaps, 141, 23

\bibitem[{{Rafert} \& {Twigg}(1980)}]{Rafert-1980}
{Rafert}, J.~B. \& {Twigg}, L.~W. 1980, \mnras, 193, 79

\bibitem[{{Reimers}(1975)}]{Reimers-1975}
{Reimers}, D. 1975, Mem. Soc. Roy. Sci. Li\`ege, 6th Ser., 8

\bibitem[{{Ritter}(1996)}]{Ritter-1996}
{Ritter}, H. 1996, in Evolutionary Processes in Binary Stars, ed. R.~A.~M.~J.
  {Wijers}, M.~B. {Davies}, \& C.~A. {Tout} (Dordrecht: Kluwer), 223

\bibitem[{{Sandin}(2008)}]{Sandin-2008}
{Sandin}, C. 2008, \mnras, 385, 215

\bibitem[{{Sandin} \& {H{\"o}fner}(2003)}]{Sandin-2003}
{Sandin}, C. \& {H{\"o}fner}, S. 2003, \aap, 398, 253

\bibitem[{{Schatzman} {et~al.}(1993){Schatzman}, {Praderie}, \&
  {King}}]{Schatzman-93}
{Schatzman}, E.~L., {Praderie}, F., \& {King}, A.~R. 1993, {The Stars} (The
  Stars.~Schatzman, Evry L., Praderie, Francoise, pp.~402.~ISBN
  3-540-54196-9.~Springer-Verlag Berlin Heidelberg 1993)

\bibitem[{{Schr{\"o}der} {et~al.}(2003){Schr{\"o}der}, {Wachter}, \&
  {Winters}}]{Schroder-03}
{Schr{\"o}der}, K.-P., {Wachter}, A., \& {Winters}, J.~M. 2003, \aap, 398, 229

\bibitem[{{Schuerman}(1972)}]{Schuerman-1972}
{Schuerman}, D.~W. 1972, Ap\&SS, 19, 351

\bibitem[{{Shimada} {et~al.}(1994){Shimada}, {Ito}, {Hirata}, \&
  {Horaguchi}}]{Shimada-1994}
{Shimada}, M.~R., {Ito}, M., {Hirata}, B., \& {Horaguchi}, T. 1994, in IAU
  Symposium, Vol. 162, Pulsation; Rotation; and Mass Loss in Early-Type Stars,
  ed. L.~A. {Balona}, H.~F. {Henrichs}, \& J.~M. {Le Contel}, 487

\bibitem[{{Soberman} {et~al.}(1997){Soberman}, {Phinney}, \& {van den
  Heuvel}}]{Soberman-1997}
{Soberman}, G.~E., {Phinney}, E.~S., \& {van den Heuvel}, E.~P.~J. 1997, A\&A,
  327, 620

\bibitem[{{Sytov} {et~al.}(2009){Sytov}, {Bisikalo}, {Kaigorodov}, \&
  {Boyarchuk}}]{Sytov-2009}
{Sytov}, A.~Y., {Bisikalo}, D., {Kaigorodov}, P., \& {Boyarchuk}, A. 2009,
  Astronomy Reports, 53, 223

\bibitem[{{Theuns} {et~al.}(1996){Theuns}, {Boffin}, \&
  {Jorissen}}]{Theuns-1996}
{Theuns}, T., {Boffin}, H. M.~J., \& {Jorissen}, A. 1996, MNRAS, 280, 1264

\bibitem[{{Theuns} \& {Jorissen}(1993)}]{Theuns-Jorissen-93}
{Theuns}, T. \& {Jorissen}, A. 1993, MNRAS, 265, 946

\bibitem[{{Tout} \& {Eggleton}(1988)}]{Tout-Eggleton-88}
{Tout}, C.~A. \& {Eggleton}, P.~P. 1988, MNRAS, 231, 823

\bibitem[{{Vanbeveren}(1977)}]{Vanbeveren-1977}
{Vanbeveren}, D. 1977, \aap, 54, 877

\bibitem[{{Vanbeveren}(1978)}]{Vanbeveren-1978}
{Vanbeveren}, D. 1978, \apss, 57, 41

\bibitem[{{Wachter} {et~al.}(2002){Wachter}, {Schr{\"o}der}, {Winters},
  {Arndt}, \& {Sedlmayr}}]{Wachter-02}
{Wachter}, A., {Schr{\"o}der}, K.-P., {Winters}, J.~M., {Arndt}, T.~U., \&
  {Sedlmayr}, E. 2002, \aap, 384, 452

\bibitem[{{Willson}(2000)}]{Willson-00}
{Willson}, L.~A. 2000, \araa, 38, 573

\bibitem[{{Zhou} \& {Leung}(1988)}]{Zhou-1988}
{Zhou}, H.-N. \& {Leung}, K.-C. 1988, \apss, 141, 257

\end{thebibliography}

\end{document}